\documentclass[12pt,preprint]{aastex}
\usepackage{graphicx}
\def\spitzer{{\it Spitzer}\space}
\def\h2{H$_2$\space}

\def\micron{$\mu$m}
\def\microns{$\mu$m\space}
\def\arcsec{$^{\prime\prime}$}
\def\arcsecs{$^{\prime\prime}$\space}
\def\arcmins{$^{\prime}$\space}

\def\deg{$^{\circ}$}
\def\degs{$^{\circ}$\space}

\slugcomment{To appear in ApJ: January 20, 2014, V781--1  }
\shorttitle{ HiRes  \spitzer  images of 89 Protostellar jets and outflows}
\shortauthors{Velusamy et al.}

\begin{document}

\title{HiRes Deconvolved \spitzer Images of 89 Protostellar Jets and Outflows: New Data on Evolution of Outflow Morphology$^{\dag}$\let\thefootnote\relax\footnotetext{$^{\dag}$For  fits images contact the lead author} }
\author{T. Velusamy\altaffilmark{1}, W. D. Langer\altaffilmark{1}, T. Thompson\altaffilmark{2} }

\altaffiltext{1}{Jet Propulsion Laboratory, California Institute of
Technology, 4800 Oak Grove Drive, Pasadena, CA 91109;
velusamy@jpl.nasa.gov,  William.D.Langer@jpl.nasa.gov }
\altaffiltext{2}{1947C East Huntington Drive, Duarte, CA 91010; timthompson3@verizon.net}


\begin{abstract}
  To study the role of protosellar  jets and outflows  in the time evolution of the parent cores and the protostars,   the astronomical community needs a large enough data base of infrared images of protostars at the highest spatial resolution possible, to reveal the details of their morphology.    \spitzer provides unprecedented sensitivity in the infrared to study both the  jet and outflow features, however its spatial resolution is limited by its 0.85m mirror.  Here we use  a  high resolution deconvolution algorithm,  ``HiRes'',  to improve the visualization of spatial morphology by enhancing resolution (to sub-arcsecond levels in the IRAC bands) and removing the
contaminating sidelobes from bright sources in a sample of  89 protostellar objects.   These  reprocessed images are useful to  detect: (i) wide angle outflow seen in scattered light; (ii) morphological details of \h2 emission in jets and  bow shocks; and (iii)  compact  features in MIPS 24 \microns  images as protostar/disk and atomic/ionic line emissions  associated with the  jets.     The HiRes  fits image data of such a large homogeneous sample presented here will be useful to the community in studying these protostellar objects.       To illustrate the utility of this HiRes sample, we show how the opening angle of  the wide angle outflows in 31 sources, all observed in the HiRes processed \spitzer images, correlates with age.  Our data suggest a power law fit to opening angle  {\it versus} age with an  exponent of $\sim$ 0.32 and 0.02, respectively for ages $\leq$ 8000 yr and $\geq$ 8000 yr.

\end{abstract}

\pagebreak
\section{Introduction}
Protostellar outflows and jets are believed to play a crucial role in determining the mass of the central protostar and its planet forming disk by
virtue of their ability to transport energy, mass, and momentum of the surrounding material, and thus terminate the infall stage in star
and disk formation. It is now well recognized that jets and outflows are essential and inherent to  the star formation process and play a key role in the structure and evolution of the molecular clouds  in star forming regions.    The protostellar  outflows  are broadly classified into two types: molecular outflows traced mainly with CO emission, and jets observed by optical line emission (cf. recent reviews by Arce et al. 2007; Bally 2007; Pudritz et al. 2007).  Observations indicate that around some protostars, high-velocity jets with a narrow opening angle are enclosed by a low-velocity outflow with a wide opening angle (e.g.  Mundt \& Fried 1983;  Velusamy et al. 2007 \& 2011) however, in others only one component is observed. Poorly collimated flows can be due to extreme precession of the jet (Shepherd et al. 2000) and indistinguishable from wide angle outflows.  The wide angle outflows, which are also observed in the scattered light, are often missed in the optical or near-IR due to instrument insensitivity to faint emission   and confusion from the bright protostellar emissions.   \spitzer provides unprecedented sensitivity in the infrared to detect the jet/outflow features.    The IRAC, IRS, and MIPS observations can provide new insight into the structure, morphology, and physical and chemical characteristics of the outflow sources as demonstrated, for example, in the observations of HH46/47  by Noriega-Crespo et al. (2004a) \& Velusamy et al. (2007),   Cep E by Noriega-Crespo et al. (2004b) \& Velusamy et al. (2011),  L1512F by Bourke et al. (2006), L1448 by Tobin et al. (2007) \& Dionatos et al. (2009),    L1251 by Lee et al. (2010), and  HH211 by Dionatos et al. (2010).  Using the IRAC colors alone, Ybarra \& Lada (2009) demonstrated the feasibility to study the pure rotational \h2 line emissions in the shocks without the need for spectroscopic data. To get the maximum information out of the Spitzer images Backus et al. (2005) and Velusamy et al. (2008) developed a deconvolution program HiRes and demonstrated that resolution enhanced reprocessing of Spitzer images (Velusamy et al. 2007;  2011)  brought  out more clearly the morphologies of: (a)  wide-angle outflow cavities in the IRAC 3.6 \& 4.5 \microns images by scattered starlight, (b) jets and bow shocks in \h2 molecular emission within the IRAC bands, and (c) the hottest atomic/ionic gas in the jet head  characterized by [Fe II]  and [S I] line emissions within the MIPS 24 \microns band.  However, to make further progress we need a large sample of objects covering protostars of different mass and luminosity and evolutionary state.

In this paper we present the deconvolved \spitzer IRAC and MIPS images of a large sample of   89  protostellar objects. These  HiRes deconvolved images  and fits image data were developed under NASA's Astrophysics and Data Analysis Program for the purpose of making them available to the astronomical community for further studies of these protostellar objects. The use of resolution enhanced (HiRes deconvolved)  \spitzer images to study the morphology and properties of the protostar and its jet-outflow components has been illustrated in our earlier papers (Velusamy et al. 2007; 2011).  It is not practical to discuss all the morphological details for   the large sample of protostars, outflows, and jets presented here;  instead, to highlight the rich detail in these HiRes images we discuss the morphology of one outflow in L1527 to reinforce the fidelty of HiRes processing and overall benefits.  In addition, to illustrate how such a large data base can be useful,   we use the opening angle of the wide angle outflows in our sample to discuss the possible evolution of opening angle with age.
    One of the outstanding issues in protostellar evolution is the role played by outflows in regulating the protostellar mass accreting process.    Outflows are effective in clearing the material from the core which feed the growth of the protostar (Arce \& Sargent 2006) and a widening of the outflow cavity with age can lead to termination of the infall (Velusamy \& Langer 1998).  The large data set here show further evidence for the opening angle to widen with age.

\section{\spitzer Data and Analysis}
It can be difficult to trace all the protostellar components, including the extended low surface brightness features associated with the outflow or jets, in the \spitzer mosaic images  because these may be   confused by the presence of side-lobes (Airy rings) surrounding  the  high brightness  protostar and/or the disk.   However by applying the HiRes algorithm one can minimize and at best remove the diffraction effects  of the brightest features, thus  enabling improved visualization of the low surface features around them. Furthermore the resolution
enhancement provides a sharper view of the outflow cavity walls, molecular jets, and bow shocks.

\subsection{Selected Jets and Outflow Sources}
The sample of  Class 0 protostars,   \h2 jets, and  outflow sources we selected for HiRes deconvolution of Spitzer images are listed in Table 1. The majority of our target protostellar objects were selected from   ``The Youngest Protostars'' Webpage   hosted by the University of Kent (http://astro.kent.ac.uk/protostars/old/) which are based on the young Class 0 objects compiled by Froebrich (2005). In addition to these objects our sample includes  some Herbig-Haro (HH) sources and a few well known jet--outflow sources. Our sample also includes one  high--mass protostar (IRAS 20126+4104; cf.  Caratti o Garatti et al. 2008) to demonstrate the use of HiRes for such sources.  Our choice for target selection was primarily based on the availability of \spitzer images in IRAC and MIPS bands in the archives and the feasibility  for reprocessing  based on the published \spitzer images wherever available.

\subsection{HiRes Deconvolution}
 To maximize the scientific return of  the \spitzer images we  use the HiRes deconvolution processing technique that makes optimal use of the spatial information in the observations.    The algorithm, ``HiRes'' and its implementation has been discussed by Backus et al. (2005) and its performance on a variety of astrophysical sources observed by {\it \spitzer} is presented by Velusamy et al. (2008).  The HiRes deconvolution algorithm is based on the Richardson-Lucy algorithm (Richardson 1972; Lucy 1974), and the Maximum Correlation Method employed by Aumann et al. (1990) for IRAS data.  As demonstrated by Velusamy et al. (2007; 2008)  the HiRes deconvolution on {\it \spitzer} images retains a high fidelity   and preserves  all the main features.  HiRes deconvolution improves the visualization of spatial morphology by enhancing resolution and removing the contaminating sidelobes from bright sources.  The benefits of HiRes include: (a) enhanced resolution of  $\sim$ 0.6\arcsecs -- 0.8\arcsecs for IRAC bands, $\sim$ 1.8\arcsecs and $\sim$7\arcsecs for MIPS 24 \microns and 70 \microns images, respectively;    (b) the ability to detect sources below the diffraction-limited confusion level; (c) the ability to separate blended sources, and thereby provide guidance to point-source extraction procedures; and, (d) an improved ability to show the spatial morphology of resolved sources. We reprocessed the images at all the IRAC bands and the MIPS 24 and 70\microns bands using all available map data in these bands in the \spitzer archives containing the selected prostellar objects.

We used the pipeline processed basic calibrated data (BCD)  and post-BCD (pBCD) down-loaded from the {\it \spitzer} Science Center (SSC) archives.
   Our nominal data processing produced reprocessed images at each band by applying the HiRes deconvolution  on BCDs following the steps outlined by Velusamy et al. (2008).     The images  containing 44 objects, as noted by ``bcd'' in column 7 of Table 1 were  processed  in this mode using BCDs as input to HiRes.   All the HiRes images using the BCD images as input  were obtained after 50 iterations. Though HiRes was originally designed to take  BCD images as input,  we have found it worked equally well when mosaic images were used (c.f.\, IRAS20293+3952 and IRAS05358+3843 by Kumar et al. 2010; M51 by Dumas et al. 2011).  For 41 protostellar objects processed after 2012 we used the post-BCD  images as input to  HiRes deconvolution taking advantage of the improved data products in the \spitzer archives.     These objects are identified in Table 1 (column 7 as ``pbcd'').  The post-BCD data are mosaic images made using the BCD images after some re-sampling applying convolutions.  Therefore in comparison to the   BCD images as input, the mosaic images represent  already  smoothed input images.  Thus the HiRes deconvolution of such images converge more slowly, and typically requires about 150 iterations to match the same level as with BCDs as input.

The overall performance of HiRes deconvolution on the \spitzer images presented here is in good agreement with those in the previously published examples (c.f. Velusamy et al. 2008).   The HiRes deconvolution interpolates well across the ``bad'' (NANed) pixels in the input images  such as due to  muxbleed, and saturation.   All bands are relatively free of artifacts
except for the IRAC channels 3 and 4, (in particular at 8 \micron)  which in some cases show streaks due to muxstripping  and a spurious secondary component adjacent to very bright sources due to uncorrected muxbleed.  Such spurious secondary components seldom occur with HiRes in the BCD input mode.  However, in the post-BCD input mode our algorithm can miss muxbleed pixels  because of the convolutions used in the post-BCDs.  Such spurious components if present in the maps are identified in  Fig. 3.1(b) to 3.53(b)     A note of caution, when interpreting multiple components in the vicinity of bright objects in 8 \microns images one   should examine more closely the sources to check whether they might be artifacts in the data or are real.

 The angular resolutions in the deconvolved images are typically in the range of 0.6\arcsecs to 0.8\arcsecs in IRAC bands 3.6 \microns \& 4.5 \micron; 0.8\arcsecs to 1\arcsecs in IRAC 5.8 \microns \& 8 \micron; $\sim$ 2\arcsecs and $\sim$ 6\arcsecs in MIPS 24 \microns and 70 \microns respectively which are a factor of 2 better than those in the mosaic images.  The resolution enhancement achieved depends on the signal-to-noise-ratio, the image coverage (redundant sets of BCD or post-BCD images used as input) and the background levels for the point sources.  The angular resolution in each HiRes image   can be obtained by examining Gaussian fits to the ``point'' sources in each image.

\section{RESULTS AND DISCUSSION}

The advantages of the enhancement in the HiRes images  to visualize the morphological details over the mosaic images are evident in all IRAC and MIPS  bands as shown in our earlier papers (Velusamy et al. 2007; 2008; 2011). In Figures 1 and 2 we show examples of   comparison between the mosaic and HiRes images to  highlight the merits of HiRes deconvolution.    In Figure 1  we reproduce one of the unprocessed 3-color images of \spitzer IRAC bands from the literature (Tobin et al. 2008) in the upper panel  and show  the corresponding image obtained with HiRes processing  in the lower panel. The resolution enhancement is evident from the sizes of the point sources in the field and its manifestation in tracing the outflow cavity walls  and the protostar environment are obvious. Figure 2 demonstrates the fidelity of the HiRes processing.  Here we use the high resolution Gemini L$^\prime$ - band (3.8 \micron)  image (with a pixel scale of 0.049\arcsec) of the outflow in L1527, observed by Tobin et al. (2010) as a ``truth image'' for comparison with the sub-arcsec resolution IRAC 3.6 \microns image obtained after the deconvolution. Note the vertices of the outflow lobes which  are  fully resolved in the Gemini L$^\prime$ - band image are remarkably consistent with  the elongated structure in the \spitzer  HiRes  image.  (The differences in the asymmetry of the brightness in the vertices may be a consequence of the scattering across the different band shapes.)

In this paper, we present in   Figures 3.1 to 3.53  the results of HiRes deconvolution of \spitzer images  containing   89 protostellar objects and associated jets and outflows.  A list of the  objects and maps are summarized in Table 1. These objects have been studied by several authors. The reference(s) in column 6 are  provided only as examples where  the basic data on the object is given.  In column 7 we list the type of the  input data used for HiRes processing as discussed in Section 2.2.  The last column (8) lists the Figure number in which  the images of the  object is shown.   Each  HiRes image typically covers  an angular size of $\sim$ 5\arcmins   and may include one or more protostellar objects or jet/outflow features  as listed in Table 1 and these are identified in the respective figures   Figures 3.1 to 3.53.  The HiRes fits data will be available in the IRSA web site in early 2014.  Our objective here is limited to  presenting an overview of the deconvolved images to give a  perspective on how this  data could support further  studies of protostellar jets and outflows.   Therefore, for illustrative purposes we only discuss one image in detail.

\subsection{HH 1-2: as an example of HiRes images}
 Fig 3.1 is an example of how our HiRes images are displayed in    Figures 3.1 to 3.53.  In each figure there are two panels: (a)  the  24 \microns emission  is overlaid as contours on the 4.5 \microns image, and (b) a RGB 3-color representation of IRAC 3.6 \microns (blue), 4.5 \microns (green) and 8 \microns (red).  We chose to display a 4.5 \microns image since it traces equally well both the scattered light and \h2 line emissions in the jet-outflow. The 24 \microns emission contours identify  the driving source (protostar/disk) associated with the jets and outflows. Furthermore, in  cases where the atomic jet features, traced by the atomic and/or ionic line emissions within the MIPS 24 \microns band, are present they are identified by the 24 \microns  contours overlaid on 4.5 \microns gray scale image.   To bring out such features more clearly, where it is useful, we show a blowup of the selected regions (indicated by the boxes in the left panel) on the right hand side of the figure.   The 3-color representation in the lower panel (labeled (b)) helps to bring out the color differences among various features, especially the wide angle outflow traced by blue representing the scattered light (predominantly at 3.6 and 4.5 \micron) and the \h2 jets/bow shocks by green and red.  The protostellar object and the jet-outflow features are also identified in the Figures in panels (a) and (b).

 HH1/HH2 is a well studied system (being the first detected Herbig-Haro object),  see, for example,  Noriega-Crespo \& Raga (2012) for detailed anatomy of jets and counter-jets as seen in \spitzer images.  It is a good example to illustrate the significant features that can be traced in the HiRes \spitzer images.  Furthermore, as far as we know   this object is the only one  for which there exists a deconvolved image using other techniques. Note that the HiRes image at 4.5 \microns in Fig. 3.1a  compares well with a recently  deconvolved image (Noriega-Crespo \& Raga, 2012) obtained by using a different algorithm,  the  deconvolution software AWAIC (A {\it WISE} Astronomical Image Co-Adder), developed by the Wide Field Infrared Survey Explorer ({\it WISE}) for the creation of their Atlas images (see, e.g., Masci \& Fowler 2009).      Of particular importance in our results, as seen in the enlarged view (in the right panels in Fig 3.1a) is the association  of 24 \microns features with the optical knots in HH2 (south-east) as identified by Raga et al. (1990) and Noriega-Crespo \& Raga (2012)    as well as with the jet and counter jet.   These 24 \microns features trace the atomic and ionic jet features as discussed in the cases of HH46/HH47 and Cep E (Velusamy et al. 2007 \& 2011). Another new feature in our results in Fig 3.1b is the detection of the wide angle outflow cavity implying that the protostar HH1-2 VLA  is driving simultaneous wide angle outflow and collimated jets.

As reported in the case of Cep E (Velusamy et al. 2011) the MIPS images  have small pointing differences ($\sim$1\arcsec) with respect to the IRAC images.  This offset  is also evident in some of the maps shown Figures 3.1 to 3.53.    The difference in the positions of the point sources between the IRAC and MIPS images  is, in part, due to the fact that MIPS used a scan mirror to change the field of view (FOV) during their observations
and the mechanism itself suffered  a hysterisis effect that increased the positional uncertainty
by 0.6\arcsecs -- 1.0\arcsec. This uncertainty is a small fraction of the MIPS beam at
24 \microns in the mosaic images,   but it becomes more obvious when comparing MIPS and IRAC
HiRes images.  Note that the MIPS images shown in the figures or the fits image data are not corrected for any pointing offset with respect to the IRAC images. Therefore, in  case detailed positional matching is required, the pointing offset can be obtained by  comparing the ``point'' sources in the MIPS 24 \microns image with their counterparts in the IRAC bands using  the HiRes fits images.   The MIPS handbook gives a 1--$\sigma$ radial uncertainty of  1.4\arcsec, in comparison with
$\sim$ 0.2\arcsecs for IRAC.

 One problem with HiRes deconvolution is that it tends to resolve out very smooth, extended low surface brightness features.  Optimal performance HiRes   requires that the background emission is fully subtracted out in each BCD prior to applying the deconvolution. Furthermore, the positivity criteria implicit in the deconvolution algorithm makes it insensitive to  negative intensities.  In rare cases (as in Fig. 3.53) though the HiRes deconvolved image   is free from the sidelobe contamination from bright sources in the image, its resolution enhancement and its lack of preserving the background can make it harder  to detect extremely extended low surface brightness emissions.

\subsection{Summary of \spitzer HiRes images}

Figures 3.1 to 3.53 highlight the prominent features in the HiRes images of the protostar regions. The HiRes fits images  contain more details over the full extent of the processed maps  and will be available in the IRSA web site in early 2014.
 {\it Spitzer}'s  coverage of a broad range of   IR emissions in the IRAC and MIPS bands with high sensitivity and  photometric stability along with the IRS data in some cases,  provide sufficient information for a comprehensive modeling of the spectral energy distribution (SED)    to derive the physical characteristics and  the evolutionary stages of protostars (cf. Robitaille et al. 2007; Forbrich et al. 2010). However, by applying the HiRes re-processing we can extract even more information from the {\it \spitzer} data. As demonstrated in Figure 3.1 the sensitivity, the resolution enhancement and removal of the diffraction lobe confusion have led to visualizing and characterizing more clearly the following:
\begin{enumerate}
\item very high dynamic range  in the maps  in which  the protostar is well resolved on a sub-arcsec scale from the surrounding jets and outflows;
\item the protostar--disk itself is detectable in the IRAC bands in many cases (with relatively low obscuration) and almost always in MIPS bands;
\item wide-angle outflow cavities in the IRAC 3.6 and 4.5 \microns images are identified by the scattered photospheric starlight; although they are visible in the Mosaic standard PostBCD products, they are brought out more clearly in the HiRes images;
\item jets and bow shocks in \h2 molecular emission within the IRAC bands; and,
\item the hottest atomic and ionic gas in the jet and jet-head traced by the [Fe II]/[S I] line emissions within MIPS 24 \microns band.
\end{enumerate}

The  \spitzer IRAC and MIPS bands offer a unique resource to study a wide range of components simultaneously: protostars, protostellar disks, outflows,
protostellar envelopes and cores. At the short wavelength IRAC bands the outflow cones are observable in scattered light from the protostar
through the cavity created by the jets and outflows (cf. Tobin et al. 2007; Velusamy et al. 2007; 2011). A significant fraction of the emission in the IRAC bands is
also considered to contain emission  from the  H$_2$ rotational lines  (e.g. Noriega-Crespo et al. 2004a,b; Smith \& Rosen, 2005; Neufeld \& Yuan, 2008) and therefore is an excellent tracer of  \h2 emission in the  protostellar jets. Recently Velusamy et al. (2007; 2011)   have shown that  molecular jets and molecular gas in the bow shocks are readily identifiable   in the IRAC bands, while the hottest atomic and ionic gases in the bow shocks, are also identifiable  in the MIPS 24 \microns band which covers a few atomic and ionic  emission lines. The dust emission from the protostar in the MIPS bands is a good diagnostic of the circumstellar disks.

 With the exception of HH 1-2 MMS 3 all the other  protostars are detected in one or more bands in  the processed images. In at least three cases, BHR 71, CG 30 and CB 230 we clearly detect the binary components and associated \h2 jets and outflows. The binary components (A,B) in L1448 IRS3 (cf. Tobin et al. 2007),  IRAS 16293-2422 (cf. Takakuwa et al. 2007) and L723 (cf. Girart et al. 2009) are marginally to well resolved in the 24 \microns HiRes images.  We clearly detect 31 wide angle outflows in scattered light in the IRAC bands  (there may be others that are less obvious due to confusion with other features).  Out of these there are 12 outflows, including B5 IRS1 which is known to have pc scale HH flows (Yu et al. 1999) in which no detectable \h2 jets or bow shocks have been observed. In about 14 \h2 jet/bow-shock systems the atomic/ionic emission in the jet/jet-heads are detected in the MIPS 24\microns images. Prominent atomic jets are evident , for example, in Figs. 3.1 (HH 1-2), 3.16 (HH 211), 3.30 (HH 92), 3.31 (HH 212), 3.37 (HH 46/47) 3.42 (Serp-SMM1), and 3.52 (Cep E).

 Thirty one of the wide angle outflows, representing a large fraction of the sample, show  the high efficiency of the scattered light in the \spitzer images as a tracer of outflow cavities.  In Fig. 4 we show an image gallery    of all 31 outflows.  Though the wide angle outflows are observed by their CO emission, a comparison between  the outflow cavities traced in the \spitzer images of HH46 and in the  ALMA maps of CO emission show that the former is broader than the latter (Arce et al. 2013). In the case of CB 26 (Launhardt et al. 2008)  a narrower outflow is also detected in CO than seen in the scattered light cavity.    Arce et al. argue that the narrower outflow seen in $^{12}$CO is a consequence of opacity effects.   In Figs. 5 and 6 we show two examples of a comparison between the \spitzer scattered light cavity and the CO outflows. In the case of L 483, shown in Fig. 5, we overlay the red and blue shifted $^{12}$CO (1-0) intensity contours on the HiRes \spitzer image at 4.5 \micron. The $^{12}$CO (1-0)  maps are from our   unpublished  OVRO observations made in 1997.  The red- shifted lobe is narrower than the blue-shifted lobe and both show narrower opening angle in comparison to that in the scattered light. The narrower opening angle for the CO outflow   may be evidence for   $^{12}$CO  opacity effects close to the source. In the second example shown in Fig. 6 we compare the scattered light in the HiRes \spitzer image at 4.5 \microns of Cep E (left panel)  with the OVRO  $^{13}$CO (1-0) maps (right panel) obtained by Moro-Mart\'{i}n et al. (2001).  Clearly the $^{13}$CO outflow seems to trace the full extent of the scattered light cavity (also shown in  Fig. 3.52) as may be expected for  $^{13}$CO which has lower opacity  compared to $^{12}$CO in the previous example.  Thus compared to CO the \spitzer images offer a more powerful tool   to study morphology of wide angle outflows detected in the scattered light. In  addition to the scattered light cavities,  the \h2 molecular jets, bow shocks, and in some cases the atomic/ionic jet components when present are also detected in the \spitzer images thus providing a more complete morphology of the protostellar outflows and jets. The predominance of simultaneous presence of collimated jets and wide angle outflows in our sample (19 out of 31) is consistent with unified jet-outflow models such as presented by Machida et al. (2008) in which   the  accretion in the compact protostellar disk drives the high velocity jets and the accretion from the extended infall envelope drives the wide angle outflows.

  Our sample contains a large number of \h2 jet/bow shocks (partly due to selection of many known HH objects). The data presented here provide a comprehensive data base to study the number,  distribution and their separation from the protostar, of the \h2 knots and bow-shocks which are indicators of the episodic ejections and/or the inhomogeneities in the surrounding ISM.

\subsection{Time evolution of outflow morphology}

Protostellar jets and  outflows carve cavities and
inject energy, momentum and turbulence into the surrounding medium (e.g., Shu et al. 2000; Bally et al. 2007).  They have a profound effect on the parent core, which is the reservoir for material accreted by the forming star and thus on  the final
properties of the newborn star as well. Outflows
originate close to the surface of the forming star (e.g., Pudritz et al. 2007), and such clearing out of the gas and dust along with a widening of the outflow with time can  lead to the
 termination of the infall phase (Velusamy \& Langer 1998).  A possible evolutionary trend in the outflow morphology (widening with time) as a function of age  was seen in a sample observed in CO (Arce \& Sargent, 2006) and is predicted in the simulations of outflow evolution (Offner et al. 2011). A broadening of outflow with age was also found by  Seale \& Looney (2008) who used the scattered light in the outflow cavities in the unprocessed Spitzer images of 27 YSOs combined with their SEDs to quantify the shapes of the outflow.

 In Table 2 we list all the 31 wide angle outflows detected in our sample   (see Fig. 4a \& b) along with the opening angles measured as discussed below.   The outflow shapes are typically parabolic: with the opening angles being the broadest at the base and narrower   at larger distances from the vertex of the cavity.  As discussed in Velusamy \& Langer (1998) to study the   evolution of the outflow, the opening angle measured near the base is most relevant.  When bipolar outflow is clearly seen we use only the widest lobe for measuring the opening angle.   We do not use the average of the   opening angles of the outflow and the counter outflow    because, in addition to the evolutionary status of the outflow, its observed characteristics may vary with factors such as the viewing geometry (inclination), and/or environmental conditions of the protostellar   and circumstellar envelopes.  For the purpose of tracing the outflow evolution with age, the most critical  outflow characteristic is the opening angle near the vertex, at the start of the outflow (c.f. Velusamy \& Langer, 1998) before it is modified by any environmental effects.  Therefore we can assume that the widest opening angle in any of the lobes would be the most representative of the opening angle for correlating with age.   The opening angles are not corrected for inclination and are listed in column 2 of  Table 2.  Note that the arrows marked in Figs. 3.1(b) to 3.53(b) are meant to draw attention to the presence of wide angle outflow and they do not represent the exact opening angles.  The values of the opening angles  in Table 2 are measured between the boundaries of the widest outflow lobe in each object.

 Following the approach of Arce \& Sargent (2006) we use  T$_{bol}$ as a measure of the protostellar age.  In Table 2   we list the values for  T$_{bol}$  collected from the literature as indicated in column 3.  When a range of values are given we use their mean value for estimating the ages.  In a few cases we estimated the T$_{bol}$ values using the   SEDs available in the literature and these are noted in the Table.    The ages were estimated from T$_{bol}$ using the empirical relationship  given by Ladd et al. (1998) in Appendix Eq. C9. The ages and their uncertainty are listed in column 5 in Table 2. The uncertainty in the estimated ages corresponds to the uncertainty in the empirical fit as given by Ladd et al.  In Figure 7 we show the plot of opening angle versus protostellar age for all 31 outflows in our sample (denoted by crosses).  The uncertainty in the ages  is also indicated. The   error in the measured opening angles is small $<$ 5\deg. We also included in this plot the data from Arce \& Sargent (2006) represented by filled triangles. The scatter in this plot is   large; but a  trend in the opening angle, increasing with age is evident. For example we can fit  power laws  as shown in Fig. 4: Opening angle, $\theta$(deg) = 6.7$\times$[t(yr)]$^{0.32}$   for  ages $<$ 8000 yr   and $\theta$(deg) = 88.5$\times$[t(yr)]$^{0.02}$ for ages $>$ 7000 yr.   (For the fits we use only the data in our sample, all observed by scattered light in the processed \spitzer images).   At ages $\ge$ 8000 yr the plot suggests a slower rate for the increase of opening angle with time,  which would be consistent    with the opening angles approaching 180\degs asymptotically. The scarcity of data at higher ages is due to the fact that our sample is primarily Class 0 objects with just two Class 1 objects (B5 IRS1 \& HH46/47). It may be noted that we do not correct the opening angle for inclination.  Though a widening trend is evident in our data any   interpretation of fit to the data is   subject to the large uncertainty in their ages.

 Our results are broadly consistent with the plots in Arce \& Sargent (2006) and Seale \& Looney (2008).  However our values for the opening angles seem higher than those in their results.  The ``intensity weighted''  opening angles estimated by Seale \& Looney using azimuthal intensity in their circular cuts are also likely  to be narrower than that measured geometrically on high spatial resolution images.
  Arce \& Sargent used CO outflow data.  Due to   opacity effects the $^{12}$CO may not trace the entire extent of the outflow cavity as that traced by scattered light    as seen in Fig.5  (also see Arce et al. 2013).  Thus the data from the HiRes processed images clearly provide the most consistent set of apparent opening angles for the entire extent of the cavity.   Nevertheless, the observed trend in the opening angle versus age is subject to the large uncertainties in estimating the ages.  Reliable age estimates combined with a more detailed characterization of the morphological shape  of the cavity   should provide observational constraints  to the theories of outflow and  star formation processes.

\section{Summary}

 By combining the high sensitivity of  {\it \spitzer} images and  reprocessing with HiRes deconvolution on a large sample of protostellar objects, we show that the jet and outflow features are more easily identified than  those in the \spitzer mosaic images alone.  These features are:  (i) wide angle outflow seen in the scattered light; (ii) morphological details of  jet driven bow shocks and jet heads or knots; (iii)  compact features in 24 \microns  image identified as atomic/ionic line emission within the MIPS band coincident with the  jet features.  The maps and the fits image data  presented here can be used to study in detail these protostellar components  as demonstrated in the case of Cep E (Velusamy et al. 2011).     We can study directly the scattered light spectrum and hence the protostellar SED in deeply embedded protostars, by separating the protostellar photospheric scattered emission in the wide angle cavity from the jet emission.   The high contrast resolution enhanced images in the \spitzer HiRes sample provide a robust description  of the morphology of the wide angle outflow cavity in 31 objects. A trend for the widening of the opening angle with age  is evident in the power law fits to this data.  We can obtain  the \h2 emission line spectra  as observed in all IRAC bands for the  knots and use their IRAC colors as probes of  the temperature and density  in the jets and bowshocks. Such spectra are useful as diagnostics of the C-type shock  excitation of pure rotational transitions of \h2 and a few \h2 vibrational emissions within each IRAC band.  Detailed modeling of the individual shocks  will help retrace the history of episodic jet activity and the associated accretion onto the protostar.  Our resolution enhanced \spitzer image data on such a large sample of protostellar objects will be a resource to future studies of these objects. It is encouraging to see that in addition to our algorithm developed for \spitzer images newer ones such as the deconvolution software AWAIC (A {\it WISE} Astronomical Image Co-Adder)  are becoming available.  The results presented in this paper support the usefulness   and the need for reprocessing with deconvolution techniques  the  protostellar data obtained by high sensitivity but low spatial resolution observations in {\it Spitzer}, {\it Herschel}, and {\it WISE} surveys.

 \acknowledgments
We acknowledge Dr C. A. Beichman for suggesting the deveopment of the HiRes deconvolution tool for Spitzer images. We also thank the referee for helpful suggestions. This publication makes use of the Protostars Webpage hosted by the University of Kent.  This ADAP (ROSES 2009)-sponsored research was conducted at the Jet Propulsion Laboratory, California Institute of Technology under contract with the National Aeronautics and Space Administration. In 1997 the Owens Valley Radio Observatory millimeter array, which we used to observe CO, was supported by the National Science Foundation grant number AST-96-13717  {\copyright  2013. All rights reserved. California Institute of Technology: USA Government sponsorship acknowledged. }

\clearpage

\begin{table*}[htbp]
\caption{{\bf List of  protostars, jets and outflows in the \spitzer HiRes processed sample}}
\vspace{0.25cm}
\hspace*{-1.5cm}
\begin{tabular}{lllllccc}
 No.	&	Primary Name	&	Other/	&	RA	&	Dec.	&	Ref	&	Data$^{\dag}$	&	Fig.	\\
	&	 	&	Region	&	 (2000)	&	 (2000)	&		&	 	&	No.	\\
\hline	 		 		 		 		 		 		 		\\		 		 		 		 		 		 		 		
1	&	HH 1-2 MMS 3	&	L1641	&	05 36 18.2	&	-06 45 45.3	&	1,2	&	bcd	&	3.1	\\
2	&	HH 1-2 MMS 2	&	L1641	&	05 36 18.8	&	-06 45 25.3	&	1,2	&	bcd	&	3.1	\\
3	&	L 1641 VLA1	&	HH1-2 VLA	&	05 36 22.8	&	-06 46 07.6	&	1,2	&	bcd	&	3.1	\\
4	&	HH 144  	&	L1641	&	05 36 21.2	&	-06 46 07	&	2,3	&	bcd	&	3.1	\\
5	&	HH 147 MMS	&	IRAS 05339-0646	&	05 36 25.2	&	-06 44 39.8	&	1	&	bcd	&	3.1	\\
6	&	L 1448-IRS2	&	L1448	&	03 25 22.5	&	+30 45 06	&	1	&	bcd	&	3.2	\\
7	&	L 1448 NW	&	L1448	&	03 25 35.6	&	+30 45 34	&	1	&	bcd	&	3.3	\\
8	&	L 1448 N IRS 3B	&	L1448	&	03 25 36.3	&	+30 45 15	&	1	&	bcd	&	3.3	\\
9	&	L 1448 N IRS 3A	&	L1448	&	03 25 36.5	&	+30 45 22	&	1	&	bcd	&	3.3	\\
10	&	L 1448 C (N)	&	L1448	&	03 25 38.9	&	+30 44 06	&	1	&	bcd	&	3.3	\\
11	&	L 1448 C (S)	&	L1448	&	03 25 39.1	&	+30 43 59	&	1	&	bcd	&	3.3	\\
12	&	NGC 1333 I1	&	IRAS03255+3103	&	03 28 38.7	&	+31 13 32	&	1	&	bcd	&	3.4	\\
13	&	IRAS 03256+3055	&	NGC1333/Bolo 33	&	03 28 44.5	&	+31 05 39.7	&	1,4	&	bcd	&	3.5	\\
14	&	NGC 1333-I2 IRAS 2A	&	IRAS 03258+3104	&	03 28 55.59	&	+31 14 37.3	&	1,5	&	bcd	&	3.6	\\
15	&	NGC 1333-I2 IRAS 2B	&	NGC1333	&	03 28 57.21	&	+31 14 19.1	&	1,5	&	bcd	&	3.6	\\
16	&	HH 12	&	NGC1333 I6 	&	03 29 01	&	+31 20 21	&	1,5	&	bcd	&	3.7	\\
17	&	HRF 46	&	NGC1333	&	03 29 10.82	&	+31 18 19.5	&	1	&	bcd	&	3.8	\\
18	&	HH 6	&	NGC1333 I7	&	03 29 13	&	+31 18 41	&	5	&	bcd	&	3.8	\\
19	&	HRF 65	&	NGC1333	&	03 29 00.51	&	+31 12 00.6	&	1	&	pbcd	&	3.9	\\
20	&	HH 344 A/B	&	NGC1333	&	03 29 00.51	&	+31 13 38	&	5	&	pbcd	&	3.9	\\
21	&	HL 3-8	&	NGC1333	&	03 29 06	&	+31 12 00.6	&	9	&	pbcd	&	3.10	\\
22	&	HH 7-11 	&	NGC1333	&	03 29 03.06	&	+31 15 51.7	&	5	&	pbcd	&	3.11	\\
23	&	SVS 13 B MMS3	&	NGC1333 I13C	&	03 29 01.95	&	+31 15 38.3	&	1	&	pbcd	&	3.11 	\\
24	&	SVS 13 B MMS2	&	NGC1333 I13B	&	03 29 03.06	&	+31 15 51.7	&	1	&	pbcd	&	3.11	\\
25	&	SVS 13 B MMS1	&	NGC1333 I13	&	03 29 03.76	&	+31 16 04.0	&	1	&	pbcd	&	3.11	\\
26	&	HH 7-11 MMS6	&	NGC1333	&	03 29 04.00	&	+31 14 46.7	&	1	&	pbcd	&	3.9	\\
27	&	NGC 1333-I4 A2	&	NGC1333	&	03 29 10.42	&	+31 13 32.2	&	1,6	&	pbcd	&	3.10	\\
28	&	NGC 1333-I4 A1	&	NGC1333 	&	03 29 10.53	&	+31 13 31.1	&	1,6	&	pbcd	&	3.10	\\
29	&	NGC 1333-I4 B	&	NGC1333	&	03 29 13.6	&	+31 13 06.6	&	1,6	&	pbcd	&	3.10	\\
30	&	NGC 1333-I4C	&	NGC1333	&	03 29 13.62	&	+31 13 57.9	&	1,6	&	pbcd	&	3.10	\\
31	&	ASR 57	&	NGC1333	&	03 29 14.5	&	+31 14 44	&	6	&	pbcd	&	3.12	\\
32	&	HH 347 A/B	&	NGC1333	&	03 29 14.5	&	+31 14 44	&	5	&	pbcd	&	3.12	\\
33	&	HH  5 A/B	&	NGC1333	&	03 29 20	&	+31 12 48	&	5	&	pbcd	&	3.13	\\
34	&	IRAS 03282+3035	&	NGC1333	&	03 31 20.3	&	+30 45 25	&	1	&	pbcd	&	3.14	\\
35	&	Bolo 102	&		&	03 43 51.1	&	+32 03 23	&	1,3	&	bcd	&	3.15	\\
\hline
\end{tabular}
\end{table*}

\begin{table*}[htbp]
\renewcommand{\thetable}{1}
\caption{{\bf continued}}
\vspace{0.25cm}
\hspace*{-1.5cm}
\begin{tabular}{lllllccc}
 No.	&	Primary Name	&	Other/	&	RA	&	Dec.	&	Ref	&	Data$^{\dag}$	&	Fig.	\\
	&	 	&	Region	&	 (2000)	&	 (2000)	&		&	 	&	No.	\\
\hline	 		 		 		 		 		 		 		\\		 		 		 		 		 		 		 		
36	&	HH 211 MMS	&		&	03 43 56	&	+32 00 48.0	&	1	&	bcd	&	3.16	\\
37	&	IC 348 MMS	&		&	03:43:57.2	&	32:03:05	&	1	&	bcd	&	3.17	\\
38	&	B 5-IRS1	&	B5	&	03 47 41.6	&	+32 51 43	&	7	&	bcd	&	3.18	\\
39	&	B 213	&	IRAS04166+2706	&	04 19 42.6	&	27 13 38 	&	1	&	pbcd	&	3.19	\\
40	&	L 1551-IRS 5	&	IRAS 04287+1801	&	04 31 34.15	&	+18 08 05.2	&	1	&	bcd	&	3.20	\\
41	&	L 1551-NE A/B	&		&	04 31 44.47	&	+18 08 31.9	&	1	&	bcd	&	3.21	\\
42	&	IRAS 04325+2402	&	L 1535 IRS	&	04 35 35.0	&	+24 08 22	&	1	&	pbcd	&	3.22	\\
43	&	L 1527 	&	IRAS 04368+2557 	&	04 39 53.9	&	+26 03 11	&	1	&	bcd	&	3.23	\\
44	&	CB 26 	&	L 1429	&	04 59 50.74	&	52 04 43.8	&	9	&	pbcd	&	3.24	\\
45	&	L 1634 IRS 7	&		&	05 19 51.5	&	-05 52 06	&	1	&	pbcd	&	3.25	\\
46	&	IRAS 05173-0555	&	L 1634	&	05 19 48.9	&	-05 52 05	&	1	&	pbcd	&	3.25	\\
47	&	Haro-4-357	&	MHO 81	&	05 35 15	&	-06 13 40	&	10	&	bcd	&	3.26	\\
48	&	Haro-4-352	&	MHO 117, HH40	&	05 35 21	&	-06 18 23	&	10	&	bcd	&	3.27	\\
49	&	HH 34	&		&	05 35 30	&	-06 26 58	&	11	&	bcd	&	3.28	\\
50	&	L 1641-N	&		&	05 36 18.6	&	-06 22  10	&	1	&	bcd	&	3.29	\\
51	&	L 1641 SMS III	&		&	05 36 24.0	&	-06 24 54	&	1	&	bcd	&	3.29	\\
52	&	HH 92	&	IRAS 05399-0121	&	05 42 28	&	-01 20 01	&	7	&	bcd	&	3.30	\\
53	&	HH 212-MM	&		&	05 43 51.1	&	-01 03 01	&	1	&	bcd	&	3.31	\\
54	&	HH 25 MMS	&		&	05 46 07.8	&	-00 13 41	&	1	&	bcd	&	3.32	\\
55	&	HH 26 IR	&		&	05 46 05.4	&	-00 14 16.6	&	8	&	bcd	&	3.32	\\
56	&	HH 24 MMS	&		&	05 46 08.8	&	-00 10 47	&	1	&	bcd	&	3.33	\\
57	&	HH 111 MMS	&		&	05 51 46.3	&	+02 48 28	&	1	&	bcd	&	3.34	\\
58	&	NGC 2264 G-VLA 2	&		&	06 41 10.9	&	+09 56 02	&	1	&	bcd	&	3.35	\\
59	&	CG 30-N/ BHR12	&	IRAS 08076-3556	&	08 09 33.1	&	 -36 04 58.1	&	1,12	&	pbcd	&	3.36	\\
60	&	CG 30-S	&	IRAS 08076-3556	&	08 09 32.7	&	-36 05 19.1	&	1,12	&	pbcd	&	3.36	\\
61	&	HH 46/47	&		&	08 25 43	&	-51 00 36	&	13	&	bcd	&	3.37	\\
62	&	BHR 71 IRS1	&	IRAS 11590-6452	&	12 01 36.8	&	-65 08 49	&	1,12	&	pbcd	&	3.38	\\
63	&	BHR 71 IRS2	&	IRAS 11590-6452	&	12 01 34.1	&	-65 08 47	&	1,12	&	pbcd	&	3.38	\\
64	&	IRAS 15398-3359	&		&	15 43 01.3	&	-34 09 12	&	1	&	pbcd	&	3.39	\\
65	&	IRAS 16293-2422 A/B	&		&	16 32 22.62	&	-24 28 32.3	&	1	&	pbcd	&	3.40	\\
66	&	L 483	&	IRAS 18148-0440	&	18 17 29.8	&	-04 39 38.3	&	1	&	bcd	&	3.41	\\
67	&	Serp-FIRS1(SMM1)	&	IRAS 18273+0113	&	18 29 49.80	&	01 15 20.6 &	1,17	&	pbcd	&	3.42 	\\
68	&	Serp-S68N	&	 	&	18 29 48.1	&	01 16 41 &	1 	&	pbcd	&	3.42 	\\
69	&	Serp-SMM5	&	 	&	18 29 51.1 &	01 16 36 &	1 	&	pbcd	&	3.42 	\\
70	&	Serp-SMM10	&	 	&	18 29 52.1	&	01 15 48 &	1 	&	pbcd	&	3.42 	\\

\hline
\end{tabular}
\end{table*}

\begin{table*}[htbp]
\renewcommand{\thetable}{1}
\caption{{\bf continued}}
\vspace{0.25cm}
\hspace*{-1.5cm}
\begin{tabular}{lllllccc}
 No.	&	Primary Name	&	Other/	&	RA	&	Dec.	&	Ref	&	Data$^{\dag}$	&	Fig.	\\
	&	 	&	Region	&	 (2000)	&	 (2000)	&		&	 	&	No.	\\
\hline	 		 		 		 		 		 		 		\\	
71	&	L 723	&	HH223	&	19 17 53.16	&	+19 12 16.6	&	1	&	pbcd	&	3.43	\\
72	&	B 335	&	IRAS 19345+0727	&	19 37 01.03	&	+07 34 10.9	&	1	&	bcd	&	3.44	\\
73	&	IRAS 20126+4104	&		&	20 14 26.0	&	+41 13 32	&	14	&	pbcd	&	3.45	\\
74	&	L1152	&	IRAS 20353+6742	&	20 35 45.9	&	+67 53 02	&	1	&	pbcd	&	3.46	\\	 		 		 		 		 		 		 		 75	&	L 1157	&	IRAS 20386+6751 	&	20 39 06.5	&	+68 02 13	&	1	&	bcd	&	3.47	\\
76	&	L 1228	&		&	20 57 13.00	&	+77 35 43.6	&	1	&	bcd	&	3.48	\\
77	&	CB 230 A	&	IRAS 21169+6804	&	21 17 38.5	&	+68 17 33.0	&	1,15	&	pbcd	&	3.49	\\
78	&	CB 230 B	&	IRAS 21169+6805	&	21 17 40.3	&	+68 17 32.7	&	1,15	&	pbcd	&	3.49	\\
79	&	L 1251B IRS4	&		&	22 38 42.80	&	+75 11 36.8	&	1	&	pbcd	&	3.50	\\
80	&	L 1251B IRS1	&		&	22 38 46.9	&	+75 11 33.9	&	1	&	pbcd	&	3.50	\\
81	&	L 1251 B	&		&	22 38 47.2	&	+75 11 28.8	&	1	&	pbcd	&	3.50	\\
82	&	L 1251B IRS2	&		&	22 38 53.0	&	+75 11 23.5	&	1	&	pbcd	&	3.50	\\
83	&	L 1251B 16	&		&	22 39 13.3	&	+75 12 15.8	&	1	&	pbcd	&	3.51	\\
84	&	L 1211 MMS1	&		&	22 47 02.2	&	+62 01 32	&	1	&	pbcd	&	3.51	\\
85	&	L 1211 MMS2	&		&	22 47 07.6	&	+62 01 26	&	1	&	pbcd	&	3.51	\\
86	&	L 1211 MMS3	&		&	22 47 12.4	&	+62 01 37	&	1	&	pbcd	&	3.51	\\
87 &	L 1211 MMS4	&		&	22 47 17.2	&	+62 02 34	&	1	&	pbcd	&	3.51	\\
88	&	Cep E-MM	&		&	23 03 13.1	&	+61 42 26	&	1	&	bcd	&	3.52	\\
89	&	IRAS 23238+7401	&		&	23 25 46.4	&	+74 17 38	&	16	&	pbcd	&	3.53	\\
\hline
\end{tabular}
\\
$^{\dag}$The terms bcd and pbcd refer to the BCD and post-BCD data from \spitzer archives which were used for HiRes deconvolution (see text).\\
		References: 				
(1)		The Youngest Protostars Webpage:  http://astro.kent.ac.uk/protostars/old/;				
(2)		Noriega-crespo \& Raga   2012 ;				
(3)		Reipurth et al.  1993;				
(4)		Enoch et al.  2009;				
(5)		Bally et al.  1996;				
(6)		Choi et al. 2006;				
(7)		Connellley et al.  2008;				
(8)		Wu et al.  2004;				
(9)		Launhardt et al. 2008;				
(10)		MHO catalog:  http://www.astro.ljmu.ac.uk/MHCat/;				
(11)		Reipurth et al. 1992;				
(12)		Chen et al. 2008;				
(13)		Noriega-crespo et al. 2004;				
(14)		Caratti o Garatti et al. 2008;				
(15)		Massi et al. 2004;				
(16)		Seale \& Looney 2008;
(17)		Choi 2009 		
				
\end{table*}

\begin{table*}[htbp]
\begin{center}
\renewcommand{\thetable}{2}
\caption{{\bf Opening Angle of Wide Angle Outflow Cavities}}
\vspace{0.25cm}
\begin{tabular}{lcccc}
Primary Name  &Opening	&\multicolumn{2}{c} {	T$_{bol}$ }	&	age $^\ddag$	\\
 	& angle(deg)$^\dag$ &	(K)	&ref.	&	10$^3$ yr	\\
\hline	 		 		 		 		 		 		 		 		\\
L1641 VLA1	&	50	&	41	&	1	&	0.9	(	0.2	,	3.7	)	\\
L 1448-IRS2	&	70	&	43-63	&	2	&	2.6	(	0.7	,	10.4	)	\\
L 1448 C (N)	&	65	&	49-69	&	2	&	3.3	(	0.8	,	13.0	)	\\
HH12	&	100	&	304	&	2	&	115	(	29	,	456	)	\\
IRAS 03256+3055	&	100	&	67	&	2	&	3.0	(	0.8	,	12.1	)	\\
IRAS03282+3035	&	80	&	33-60	&	3	&	1.3	(	0.3	,	5.2	)	\\
Bolo 102	&	95	&	72	&	2	&	3.6	(	0.9	,	14.4	)	\\
HH 211-MM	&	50	&	24	&	2	&	0.3	(	0.1	,	1.0	)	\\
B5 IRS1	&	130	&	287	&	2	&	100	(	25 	,	397 	)	\\
B213	&	80	&	72	&	2	&	3.6	(	0.9	,	14.4	)	\\
L 1551-IRS 5	&	105	&	92	&	4	&	6.5	(	1.6	,	25.9	)	\\
L 1551-NE A/B	&	100	&	91	&	4	&	6.3	(	1.6	,	25.2	)	\\
IRAS 04325+2402	&	110	&	73	&	4	&	3.7	(	0.9	,	14.9	)	\\
L1527 	&	98	&	56	&	4	&	2.0	(	0.5	,	7.9	)	\\
CB26 	&	115	&	770	&	5	&	1065 	(	268	,	4242 	)	\\
L 1634 IRS7	&	76	&	64$^*$	&	6	&	2.7	(	0.7	,	10.8	)	\\
IRAS05173-0555	&	64	&	58$^*$	&	6	&	5.7	(	1.4	,	22.6	)	\\
HH92	&	60	&	41$^*$	&	7	&	0.9	(	0.2	,	3.7	)	\\
HH111MMS	&	118	&	78	&	4	&	4.4	(	1.1	,	17.4	)	\\
IRAS08076-3556	&	42	&	117	&	4	&	0.7	(	0.2	,	2.9	)	\\
HH46/47	&	110	&	40-145	&	8	&	19.4	(	4.9	,	77.1	)	\\
BHR71-IRS1	&	45	&	44	&	9	&	1.1	(	0.3	,	4.4	)	\\
BHR71-IRS2	&	42	&	58	&	9	&	2.1	(	0.5	,	8.6	)	\\
IRAS15398-3359	&	120	&	48-61	&	4	&	2.4	(	0.6	,	9.7	)	\\
L483	&	70	&	50-54	&	3	&	1.8	(	0.5	,	7.2	)	\\
B335	&	63	&	28-45	&	3	&	1.2	(	0.3	,	4.7	)	\\
L1152	&	64	&	48$^*$	&	6	&	1.4	(	0.3	,	5.4	)	\\
L1157	&	54	&	40-60	&	3	&	1.0	(	0.2	,	3.9	)	\\
CB230	&	95	&	69	&	4	&	3.3	(	0.8	,	13.0	)	\\
CepE-MM	&	85	&	56	&	4	&	2.0	(	0.5	,	7.9	)	\\
IRAS23238+7401	&	98	&	64$^*$	&	6	&	2.7	(	0.7	,	10.8	)	\\
\hline
\end{tabular}
\end{center}
$^\dag$This paper\\
$^\ddag$1--$\sigma$ lower and upper limits are given in parenthesis\\
$^{*}$Estimated from SED peak. \\
References for T$_{bol}$:  (1)	Fischer et al. 2010;
(2)	Enoch   et al. 2009;
(3)	Chen et al. 2013;
(4)	Froebrich 2005;
(5)	Stecklum et al. 2004;
(6)	Seale \& Looney 2008;
(7)	Miettinen et al.2013;
(8)	Emerson et al. 1994;
(9)	Chen et al. 2008.
\end{table*}

 \clearpage
\begin{figure*}[t]
\begin{center}
\vspace{-1.0cm}
 \includegraphics  [ angle=0, scale =0.8]{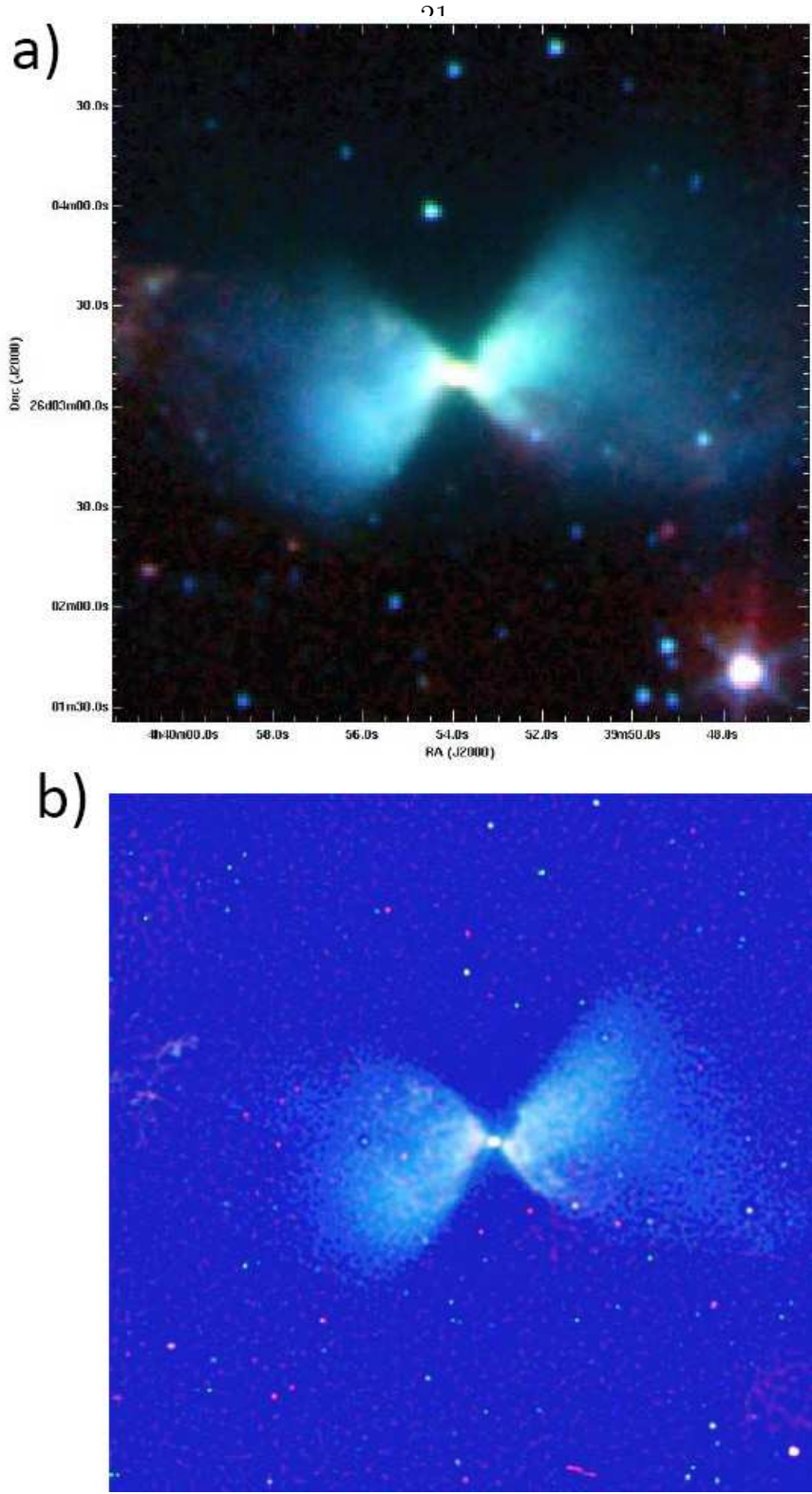}
 \caption{Example of a comparison of HiRes denconvolved \spitzer IRAC images   with  unprocessed mosaic images in  the literature:   false 3-color IRAC images of L1527 (blue: 3.6 \micron; green: 4.5 \micron; and red: 8.0 \microns). (a) unprocessed mosaic image reproduced from Tobin et al. (2008). (b)  the HiRes deconvolved  image.
 \label{demo1}}
 \end{center}
 \end{figure*}

 \begin{figure*}[t]
 \begin{center}
 \vspace{-1.0cm}
 \includegraphics  [ angle=0, scale =0.8]{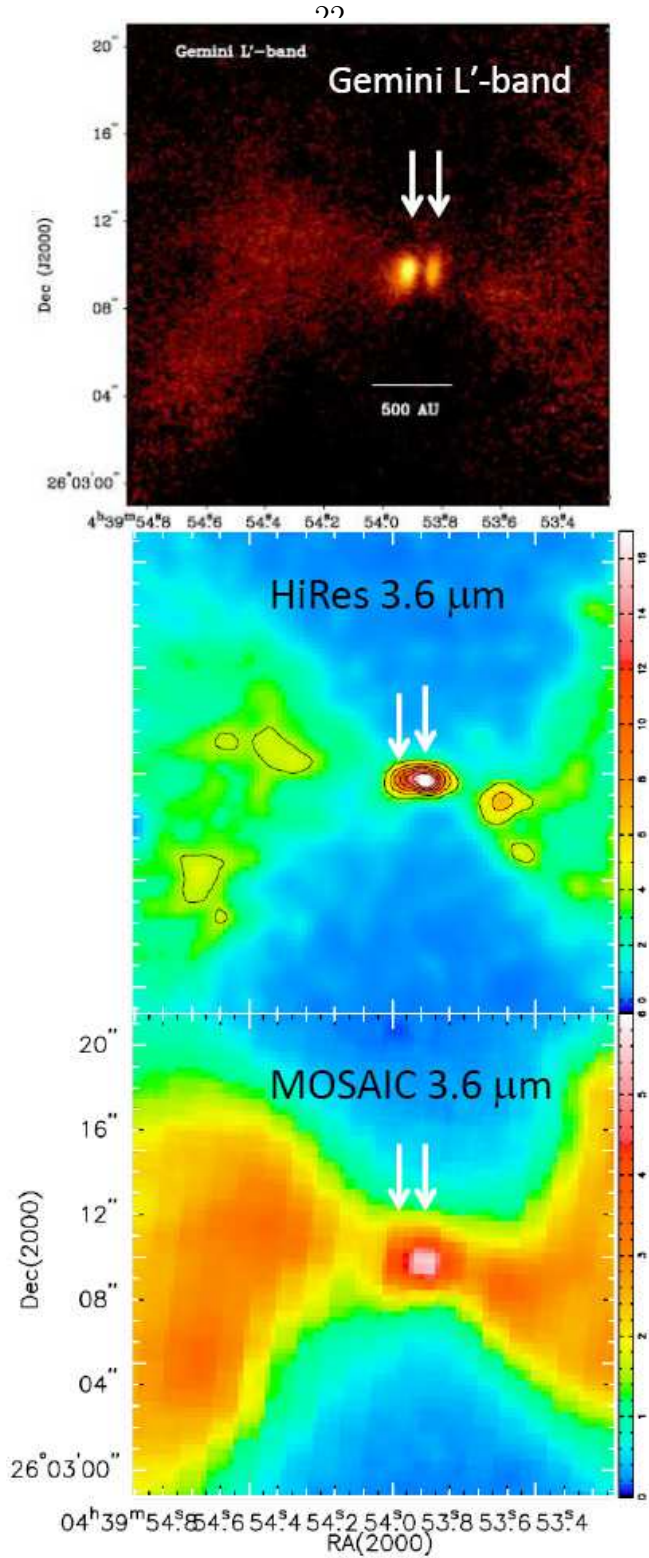}
 \caption{\spitzer 3.6 \microns HiRes deconvolved image compared with  high resolution (with a pixel scale of 0.049\arcsec) Gemini L$^\prime$ - band (3.8 \micron) image of the outflow in L1527.  For comparison the IRAC 3.6 \microns mosaic image is shown in the lower panel.  Note the sub-arcsec resolution enhancement in the deconvolved image (middle panel) brings out the innermost structure as observed in the high resolution image in the top panel  (reproduced from Tobin et al. 2010).
  \label{demo2}}
 \end{center}
 \end{figure*}
\clearpage
\setcounter{figure}{0}
\begin{figure*}[t]
\makeatletter
\vspace{-1.0cm}
\renewcommand{\thefigure}{3.1a}
\hspace*{-1.5cm}
\includegraphics  [ scale=0.72, angle=-90]{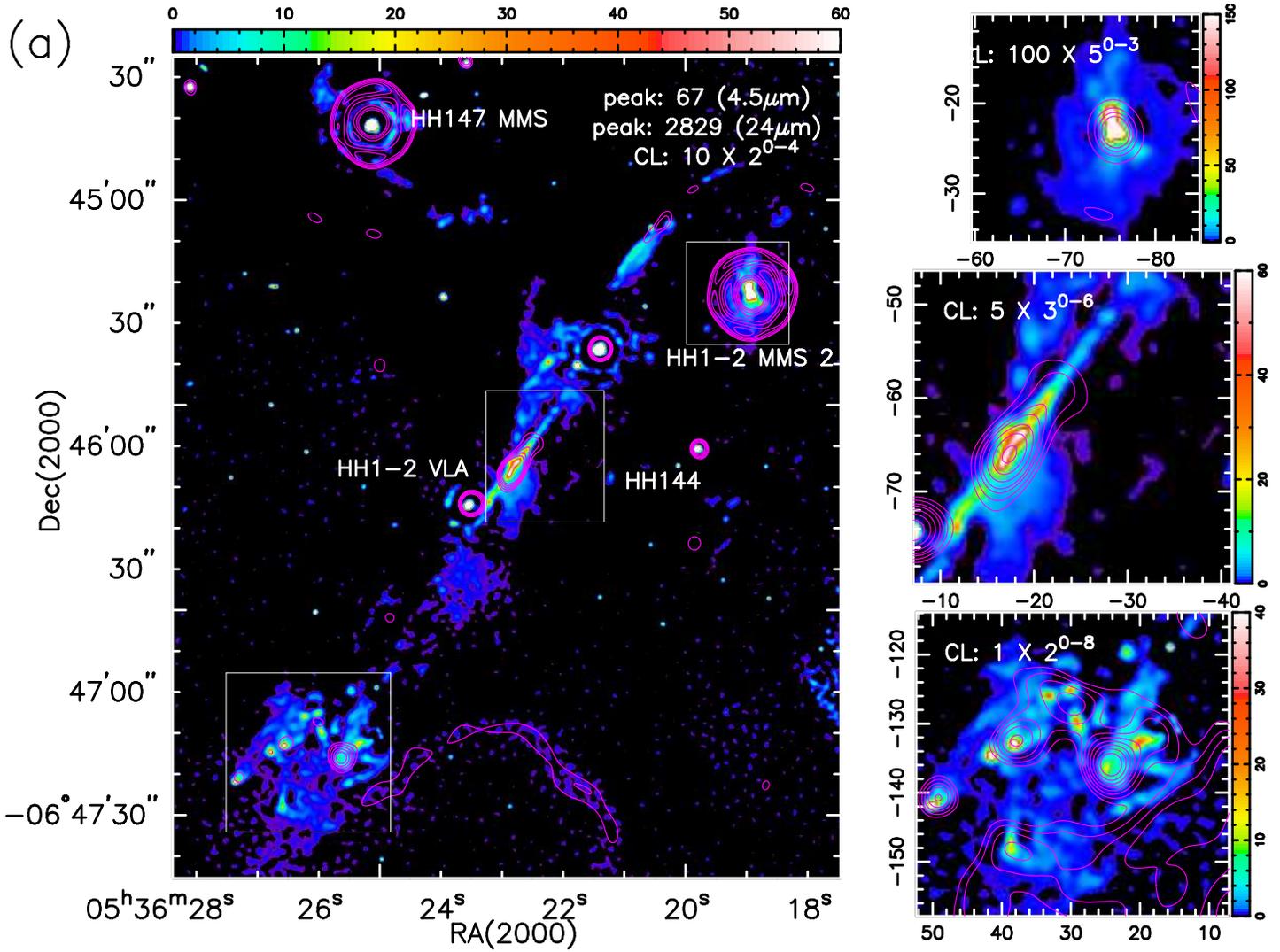}
\caption{  HH1-2:  The HiRes MIPS 24\microns image overlaid as contours on the IRAC 4.5\microns  image. For details see caption for FIGURE SET 3 below.
\label{HH1_2a}}
\end{figure*}
\clearpage
\setcounter{figure}{0}
\begin{figure*}[t]
\makeatletter
\vspace{-0.75cm}
\renewcommand{\thefigure}{3.1b}
\hspace*{-1.5cm}
\includegraphics  [ angle=0, scale =0.7]{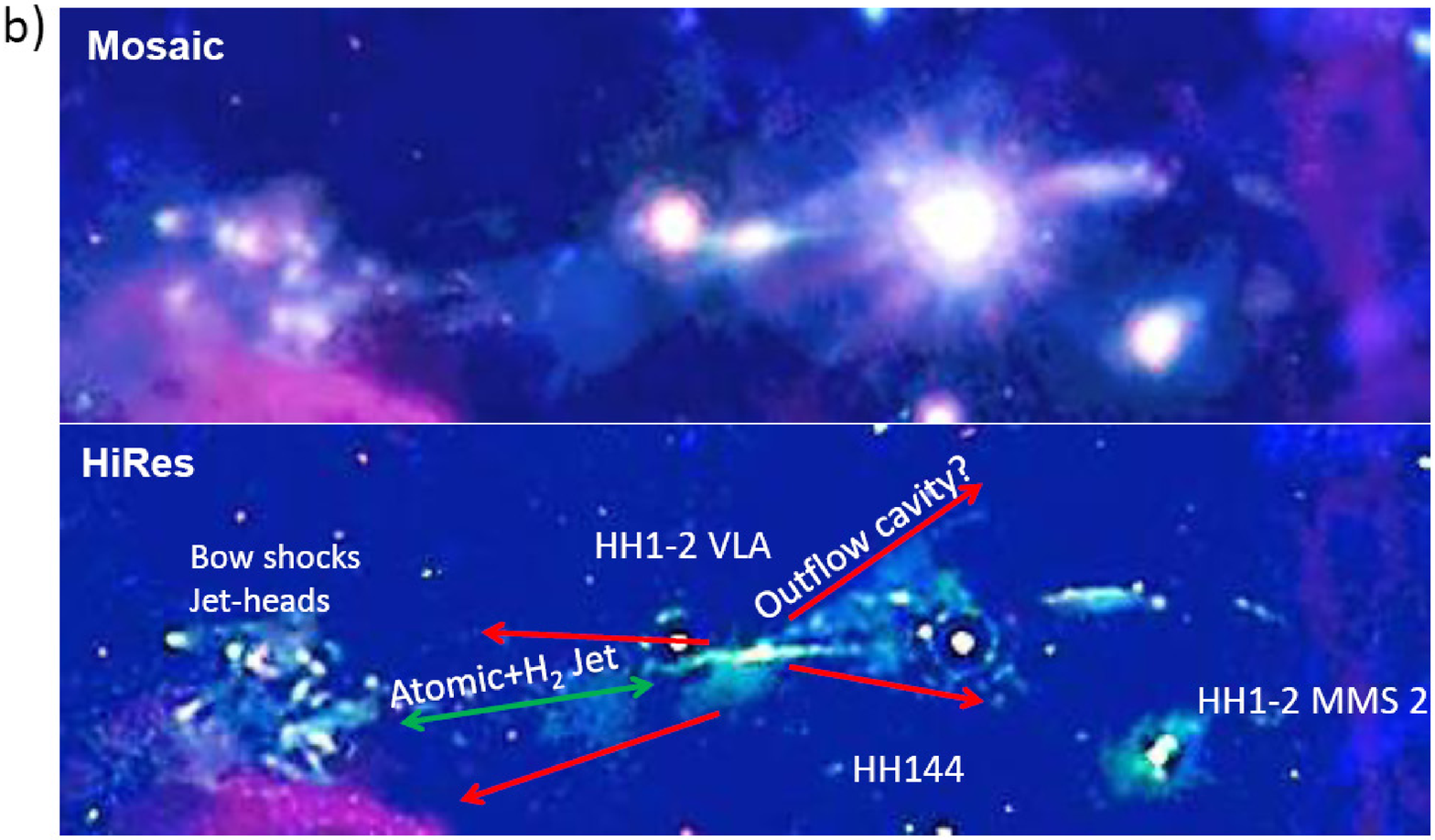}
\caption{  HH1-2:   3-color mosaic and HiRes  images: 3.6  \microns (blue), 4.5 \microns (green), 8.0 \microns (red). For details see caption for FIGURE SET 3 below. \newline
FIGURE SET 3: See the electronic edition of the Journal for all 53 images. \newline { Caption to \bf Figures 3.1 to 3.53:   (a)} The 4.5 \microns and 24 \microns HiRes deconvolved  \spitzer IRAC  image of the protostellar jet and outflow in our sample.  The MIPS 24\microns image overlaid as contours on the IRAC 4.5\microns gray scale image (in a few cases the 3.6\microns image is used and is  indicated in the image). A square root color stretch is used to bring   out low and high brightness emissions and the contour levels are in  increments of factors of two or larger  as indicated. The peak intensities (in units of MJy Sr$^{-1}$) in the protostar/jet-outflow are indicated.  The panels to the right  show  blowups of the selected regions (indicated by the boxes in the left panel)  where it is needed to bring out the features.  In some cases the positions of the resolved stellar components in the IRAC and 8.0\microns images are labeled. In  cases where the atomic jet features, traced by the atomic and/or ionic line emissions within the MIPS 24 \microns band are present they are identified by the 24 \microns  contours overlaid on 4.5 \microns gray scale image.    {\bf (b)}  3-color representation of HiRes deconvolved \spitzer IRAC images:   IRAC 3.6  \microns (blue), 4.5 \microns (green), 8.0 \microns (red).   In some cases the arrows (white) or numbers mark positions of the resolved stellar components in the IRAC 5.8 and 8.0 \microns images.  The images are re-orientated to conserve space and the orientation in the sky  is easily inferred from the image in the  upper panel. Muxbleed artifacts if present are indicated by a cross.  In the 3-color representation the wide angle outflows are identified by the blue excess while the \h2 jets and bow shocks by the green excess.  The \h2 jets (green arrows) and the wide angle cavity (red arrows) inferred from scatted light are   indicated (not to scale).
\label{HH1_2b}}
\end{figure*}

  \clearpage
 \setcounter{figure}{0}
 \begin{figure*}[t]
 \makeatletter
\vspace{-1.5cm}
\renewcommand{\thefigure}{4}
\hspace*{-1.5cm}
 \includegraphics  [ scale=0.85 ]{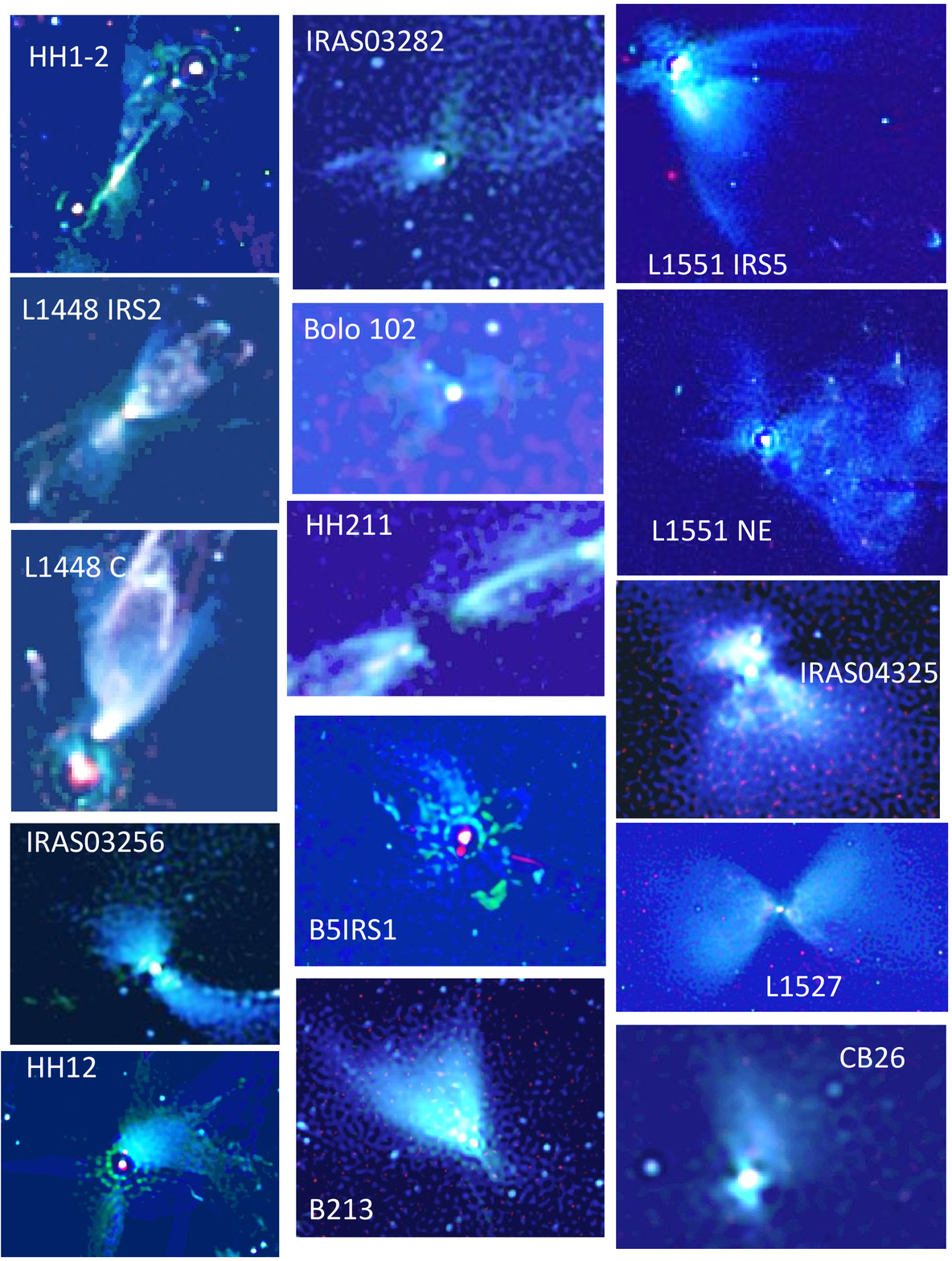}
\caption{ Image gallery of wide angle outflows observed in the HiRes processed \spitzer sample: 3-color IRAC HiRes image with 3.6  \microns (blue), 4.5 \microns (green), and 8.0 \microns (red). The wide angle cavities are identified by the color excess (blue). Note B5 IRS1 and HH12 have large side-lobe residue due to saturation.
\label{Cavities1}}
\end{figure*}
 \clearpage
 \setcounter{figure}{0}
 \begin{figure*}[t]
 \makeatletter
\vspace{-1.5cm}
\renewcommand{\thefigure}{4 }
\hspace*{-1.5cm}
 \includegraphics  [ scale=0.9 ]{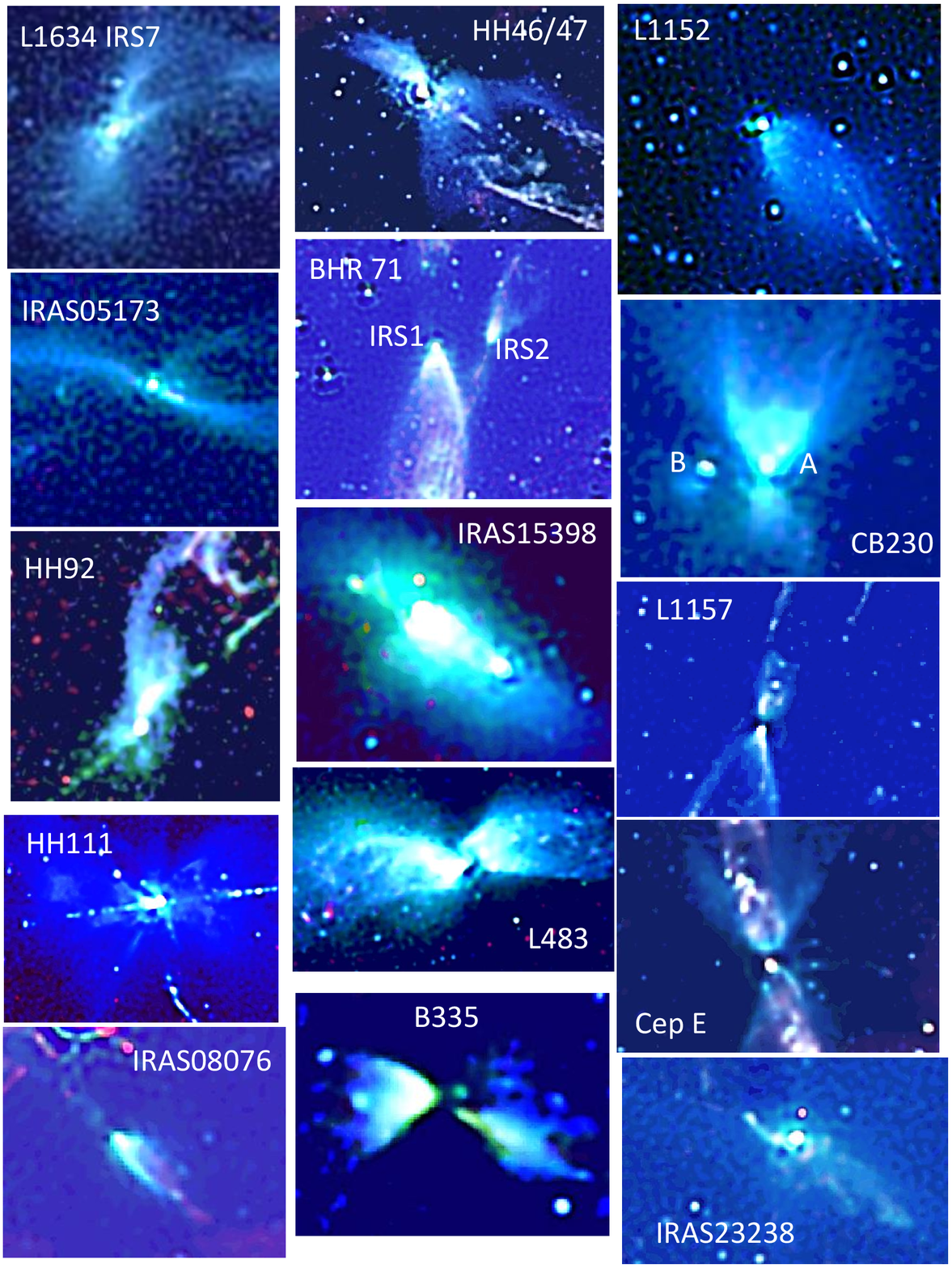}
\caption{ (Continued)
\label{Cavities2}}
\end{figure*}
 \clearpage
 \setcounter{figure}{0}
 \begin{figure*}[t]
 \makeatletter
\vspace{-1cm}
\renewcommand{\thefigure}{5}
\hspace*{-1.5cm}
 \includegraphics  [ scale=0.6]{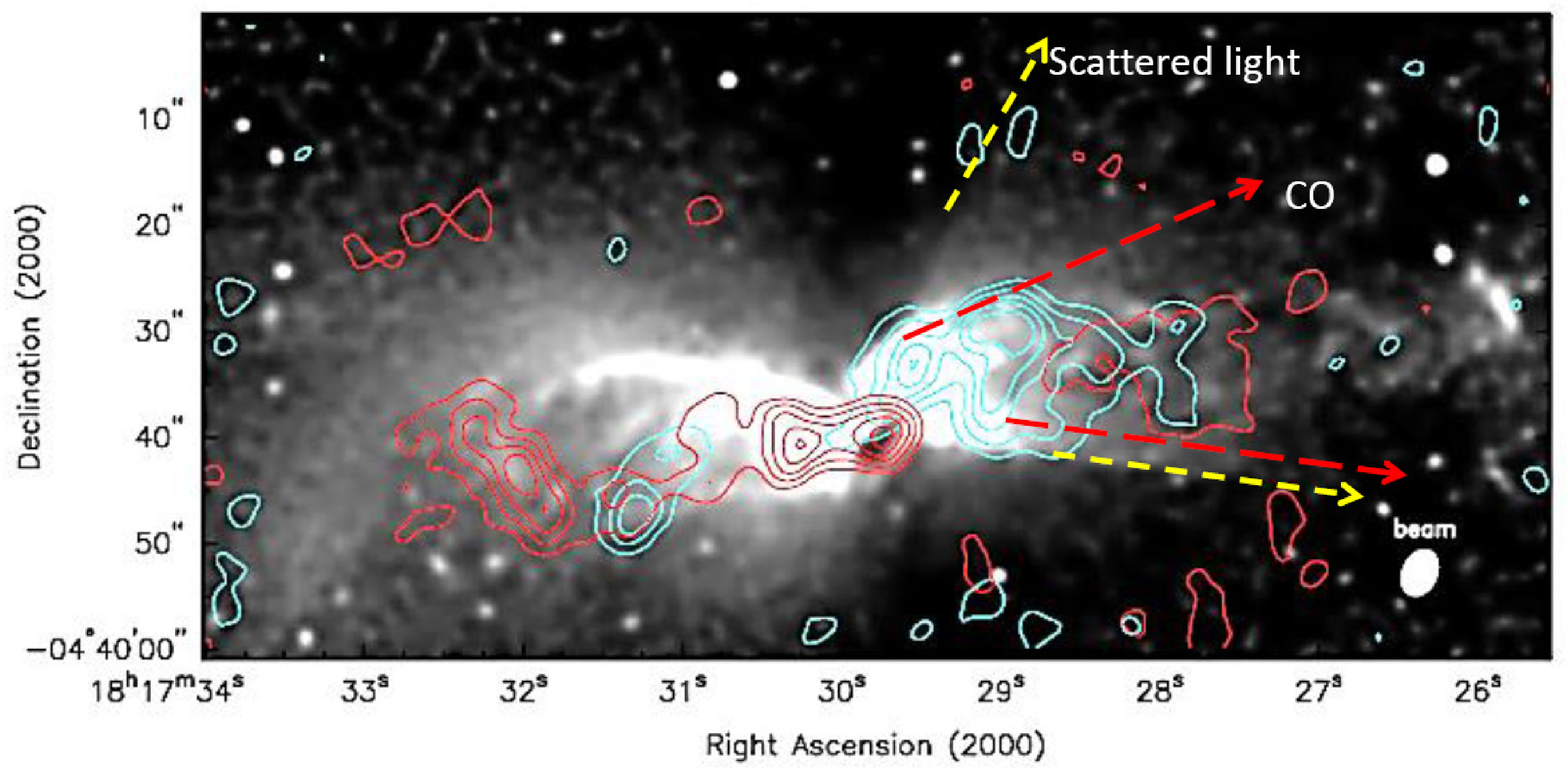}
\caption{ L483: Comparison of the scattered light outflow cavity traced in HiRes 4.5 \microns image with $^{12}$CO(1-0) outflow lobes. CO outflow intensity contours (at inteval 0.4 Jy beam$^{-1}$) overlaid on the 4.5 \microns image. Red and blue contours represent the red-shifted (V$_{lsr}$ 6.5 -- 10.4 kms$^{-1}$) and blue-shifted (V$_{lsr}$ 0.0 -- 3.9 kms$^{-1}$) lobes respectively. The rest V$_{lsr}$ is 5.2 kms$^{-1}$.  The CO data are from our OVRO observations   with a synthesized beam of 4.6\arcsecs $\times$ 3.3\arcsec. The arrows mark the outflow cavities as traced by the scattered light at 4.5 \microns and CO emissions.
\label{CO:L483}}
\end{figure*}
 \setcounter{figure}{0}
 \begin{figure*}[!h]
 \makeatletter
\vspace{-1.5cm}
\renewcommand{\thefigure}{6}
\hspace*{-1.5cm}
 \includegraphics  [ scale=0.7]{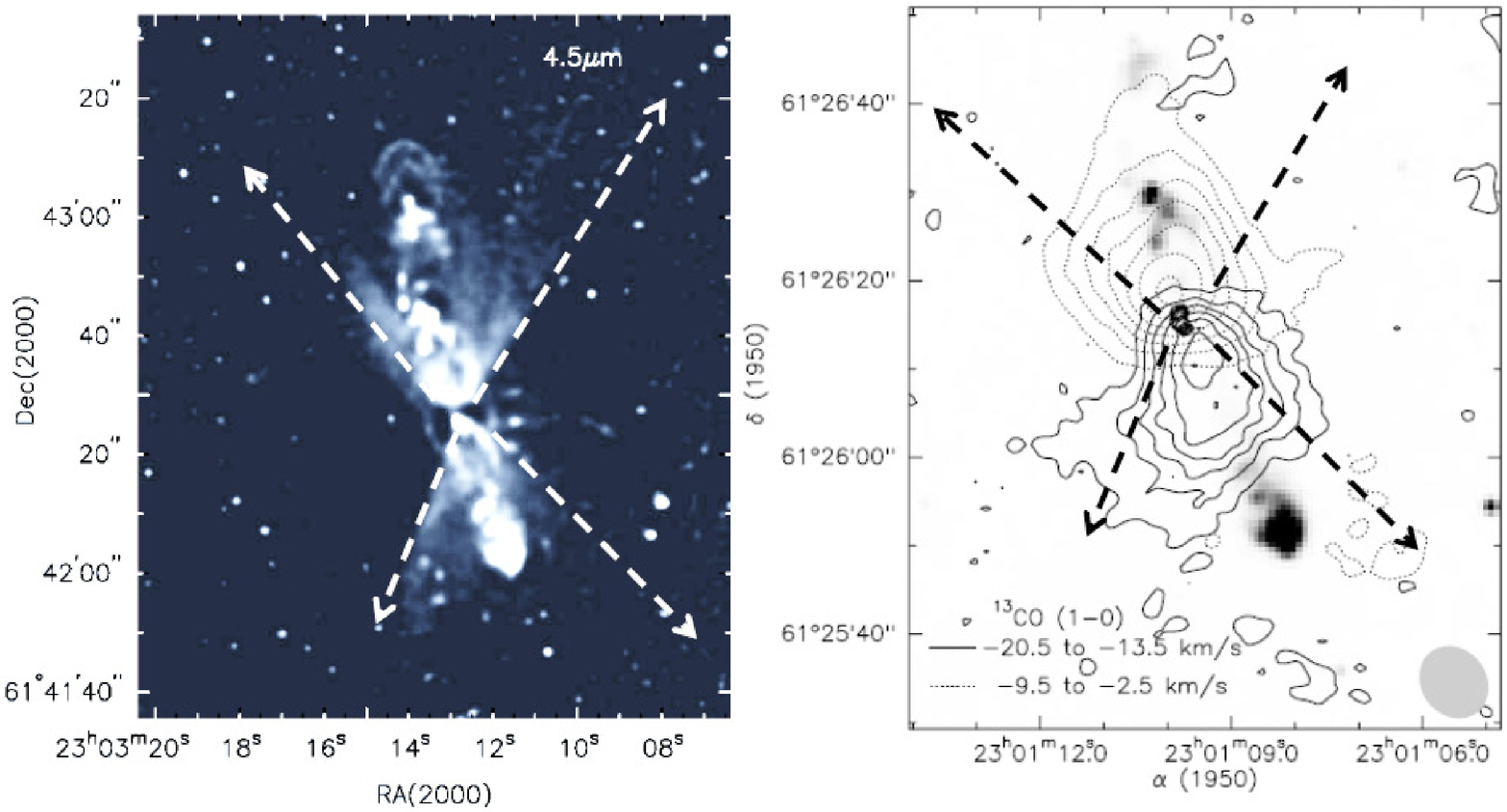}
\caption{ Cep E: Comparison of the scattered light outflow cavity traced in the HiRes 4.5 \microns image with $^{13}$CO(1-0) outflow lobes. (left) 4.5 \microns image. (right) Contour map of CO outflow intensities of    the red-shifted   and blue-shifted   lobes observed with the OVRO (reproduced from  Moro-Mart\'{i}n et al. 2001).   The arrows mark the outflow cavity traced by the scattered light at 4.5 \microns.
\label{CO:Cep E}}
\end{figure*}
\clearpage
 \setcounter{figure}{0}
 \begin{figure*}[t]
 \makeatletter
\vspace{-1.0cm}
\renewcommand{\thefigure}{7}
 \includegraphics  [ scale=0.7, angle=-90]{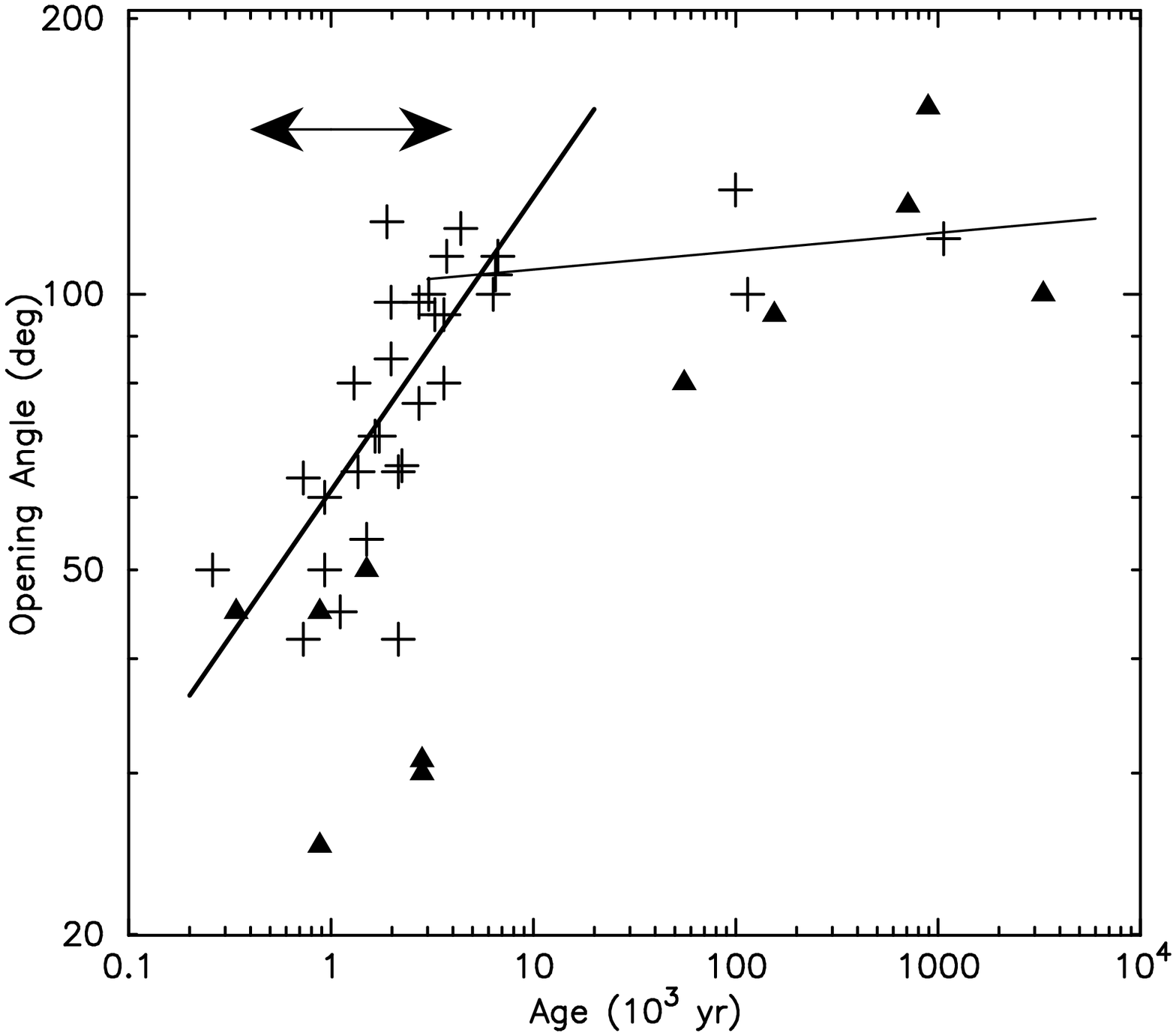}
\caption{Age versus the opening angle of all wide angle outflows detected in the HiRes images. The crosses (+) are data from the present sample, used for the power law fits. The power law fits, for opening angle versus age, are shown   for  ages $<$ 8000 yr (thick line) and  for ages $>$ 7000 yr (thin line) have exponents 0.32 and 0.02  respectively.   Filled triangles are data from Arce \& Sargent (2006) which are not used for the fit.   The double arrow indicates the uncertainty in the age estimate.
\label{Evolution}}
\end{figure*}
  \clearpage
 \begin{figure*}[t]
 \makeatletter
\vspace{-1.0cm}
\renewcommand{\thefigure}{3.2 to 3.53}
\begin{center}
\hspace*{-1.5cm}
\caption{See the electronic edition of  January 20, 2014 ApJ V781-1
\label{online Figs}}
\end{center}
\end{figure*}
\end{document}